# Photogalvanic effects in HgTe quantum wells

B. Wittmann, R. Ravash, H. Diehl, S. N. Danilov, Z. D. Kvon, S. A. Tarasenko, E. L. Ivchenko,
N. N. Mikhailov, S. A. Dvoretsky, W. Prettl, and S. D. Ganichev

*Abstract*— We report on the observation of the terahertz radiation induced circular (CPGE) and linear (LPGE) photogalvanic effects in HgTe quantum wells. The current response is well described by the phenomenological theory of CPGE and LPGE

## I. INTRODUCTION

WE report on the observation of photogalvanic effects in HgTe quantum wells (QWs). These narrow band low dimensional structures characterized by the inverted band structure and large spin splitting of subbands in the *k*-space, recently attracted growing attention as a potentially interesting material system for spintronics. Photogalvanic effects in the terahertz range has proved to be a very efficient method to study nonequilibrium processes in semiconductor QWs yielding information on their point-group symmetry, details of the band spin-splitting, processes of momentum and energy relaxation etc. [1]. In this work we investigate photogalvanic effects in this novel material as a function of the radiation polarization, wavelength, and temperature. As a result the anisotropy of the structures under study has been established and analyzed.

## II. EXPERIMENTAL TECHNIQUE AND RESULTS

The experiments are carried out on $Cd_{0.7}Hg_{0.3}Te/HgTe/Cd_{0.7}Hg_{0.3}Te$ QWs having two different widths: 16 nm and 21 nm. Structures are grown on a GaAs substrate with the surface orientation (013) by means of a modified MBE method. Samples with density of electrons $N_s$ from $1.5 \cdot 10^{11}$ to $4 \cdot 10^{11}$ cm$^{-2}$ and mobility $\mu$ at 4.2 K ranging from $2 \cdot 10^5$ to $5 \cdot 10^5$ cm$^2$/Vs are studied in the temperature range from 4.2 K up to room temperature. We note that the indicated mobility represents the highest value achieved so far in HgTe QWs for the corresponding electron density. Two pairs of contacts (along directions *x* and *y*) are centered in the middle of cleaved edges parallel to the intersection of the (013) plane and cleaved edge face {110}. For optical excitation we used high power pulsed $NH_3$, $D_2O$, and $CH_3F$ optically pumped molecular lasers as well as a Q-switched $CO_2$ laser. Linearly and circularly polarized radiation is applied in the wavelength range from 35 μm to 496 μm with power of about 20 kW and from 9.2 to 10.8 μm with power of about 1 kW. The polarization of the laser light is changed from linear to circular using quartz λ/4 plates or a ZnSe Fresnel rhombus. The helicity $P_{circ}$ of the incident light varies from –1 (left handed circular, $\sigma_-$) to +1 (right handed circular, $\sigma_+$) according to $P_{circ} = \sin 2\varphi$, where $\varphi$ is the angle between the initial plane of polarization and the optical axis of the λ/4 plate.

With illumination of samples at normal incidence we observed in the in-plane *x*-direction a current signal proportional to the helicity $P_{circ}$. The current follows the temporal structure of the laser pulse intensity and changes sign if the circular polarization is switched from left to right handed (Fig. 1). It should be stressed that the samples were unbiased, thus the irradiated samples represent a current source. The helicity dependence demonstrates that the observed current is due to CPGE. We note that, in addition to the current contribution proportional to $j \propto A \cdot P_{circ}$, we also detected a small polarization independent offset. Cooling down of the sample from room to the liquid helium temperature results in the sign inversion of the current measured at a fixed helicity. An electric current is also detected in the orthogonal *y*-direction. In this case, besides the response to the λ = 90 μm where the contribution $j \propto P_{circ}$ retains, the observed current has a more complex dependence on the angle φ. This dependence can be well fitted by

$$j = a \sin 2\varphi + b \sin 4\varphi + c \cos 4\varphi + d, \qquad (1)$$

where *a, b, c,* and *d* are fitting parameters which can be determined by solving a system of four linear equations following from Eq. (1) for angles φ = 0°, 22.5°, 45° and 135°, respectively. We have found that the ratio *A/a* between CPGE values detected in the *x*- and *y*-directions substantially changes under the variation of the radiation wavelength and temperature. The three last terms in the right-hand side of the Eq. (1) are attributed to the LPGE. This current can also be generated applying the linearly polarized radiation and has characteristic dependence on the azimuth angle α. We obtained the dependence of the current on α by rotation of the λ/2 polarizer. We found that the current in *y*-direction in most cases can well be fitted by the sin 2α dependence with a polarization independent background. In contrast, the fit in the *x*-direction, can be obtained by using a more complex trigonometric function. Experiments at oblique incidence for

Manuscript received June 17, 2007. This work was supported by the DFG via Project SPP 1285 and Collaborative Research Center SFB689.

B. Wittmann, R. Ravash, S. N. Danilov, H. Diehl, W. Prettl and S. D. Ganichev are with the Terahertz Center, University of Regensburg, 93040 Regensburg, Germany (http://www.physik.uni-regensburg.de/TerZ; phone: +499419432050; fax: +499419431657; e-mail: Sergey.Ganichev@physik.uni-regensburg.de).
Z. D. Kvon, N. N. Mikhailov, S. A. Dvoretsky are with the Institute of Semiconductor Physics, Novosibirsk, Russia.
S. A. Tarasenko and E. L. Ivchenko are with the Ioffe Physico-Technical Institute, St. Petersburg 194021, Russia (tarasenko@coherent.ioffe.ru).

the CPGE and LPGE currents show that the both currents reach a maximal value at normal incidence.

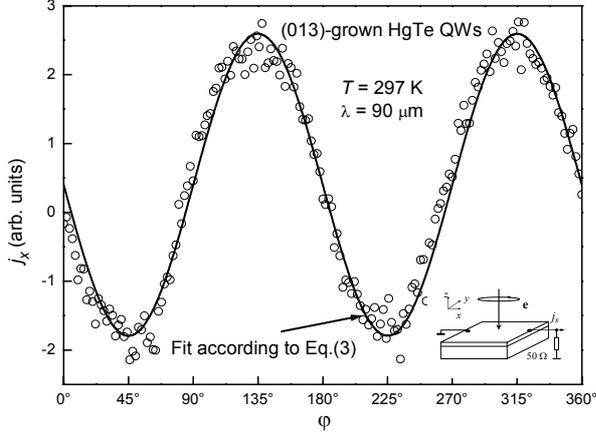

Fig. 1. Helicity dependence of the photocurrent $j_x$ in (013)-grown HgTe QW.

So far photogalvanic currents excited in zinc-blende structure based QWs by a normally incident radiation have been observed only in structures grown in [113]- and asymmetrical [110]- crystallographic directions which belong to the $C_s$ symmetry group [1]. This group contains only two symmetry elements, the identity and one mirror reflection plane $\sigma_{(1-10)}$. In this case the CPGE current flows in the direction normal to the mirror reflection plane, i.e. $x \parallel [1\bar{1}0]$. In experiments described here we use (013)-grown QWs. These structures belong to the symmetry point group $C_1$ which contains neither rotation axes nor mirror planes and consists of only one element, which is the identity operation. Phenomenologically, for the $C_1$-symmetry group, the lateral photogalvanic current for the excitation along $z \parallel [013]$ is given by

$$j_x/I = \gamma_{xz} P_{circ} + \chi_{xxx}|e_x|^2 + \chi_{xyy}|e_y|^2 + \chi_{xxy}(e_x e_y^* + e_y e_x^*), \quad (2)$$
$$j_y/I = \gamma_{yz} P_{circ} + \chi_{yxx}|e_x|^2 + \chi_{yyy}|e_y|^2 + \chi_{yxy}(e_x e_y^* + e_y e_x^*),$$

where $e$ is the radiation polarization vector, $I$ is the radiation intensity, $\gamma_{xz}$, $\gamma_{yz}$ and six linearly independent components $\chi_{xxx}$, $\chi_{xyy}$, $\chi_{xxy} = \chi_{xyx}$, $\chi_{yxy} = \chi_{yyx}$, $\chi_{yxx}$, $\chi_{yyy}$ are allowed components of the second-rank pseudo-tensor $\gamma$ describing the CPGE and the third rank tensor $\chi$ related to the LPGE. These equations yield the following polarization dependences of the photocurrent

$$j_x/I = \gamma_{xz}\sin 2\varphi + \frac{\chi_{xxx}+\chi_{xyy}}{2} + \frac{\chi_{xxx}-\chi_{xyy}}{4}(1+\cos 4\varphi) + \frac{\chi_{xxy}}{2}\sin 4\varphi \quad (3)$$

$$j_y/I = \gamma_{yz}\sin 2\varphi + \frac{\chi_{yxx}+\chi_{yyy}}{2} + \frac{\chi_{yxx}-\chi_{yyy}}{4}(1+\cos 4\varphi) + \frac{\chi_{yxy}}{2}\sin 4\varphi$$

for elliptically polarized light achieved by passing the laser radiation, initially linearly polarized along the x-axis, through a $\lambda/4$-plate, where $\varphi$ is the angle between the optical axis of the $\lambda/4$-plate and the x-axis, and

$$j_x/I = \frac{\chi_{xxx}+\chi_{xyy}}{2} + \frac{\chi_{xxx}-\chi_{xyy}}{4}\cos 2\alpha + \chi_{xxy}\sin 2\alpha, \quad (4)$$

$$j_y/I = \frac{\chi_{yxx}+\chi_{yyy}}{2} + \frac{\chi_{yxx}-\chi_{yyy}}{4}\cos 2\alpha + \chi_{yxy}\sin 2\alpha$$

for linearly polarized light, where $\alpha$ is the angle between the plane of linear polarization and the x-axis. Equations (3) and (4) well describe experimental data (see Figs. 1 and 2). The fact that the experimentally observed variation of the ratio $A/a$ with a change of the light frequency, sample temperature and radiation intensity is also in an agreement with Eqs. (3) and (4) yielding $j_x/j_y = \gamma_{xz}/\gamma_{yz}$. Indeed all above components of the tensors $\gamma$ and $\chi$ are linearly independent and the ratio between them can change with varying of experimental conditions. The direction of the circular photocurrent induced under normally incident radiation for the $C_1$ point group is not fixed in the plane of interface. The same is valid for the linear photogalvanic photocurrent, e.g., the photocurrent induced by the light polarized along the x-axis.

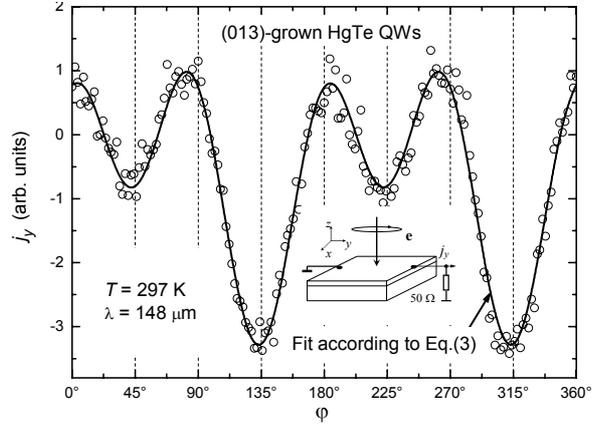

Fig. 2. Helicity dependence of the photocurrent $j_y$ in (013)-grown HgTe QW.

### III. CONCLUSION

In summary, our experiments show that photogalvanic effects can be effectively generated in HgTe quantum wells with the strength, e.g. for CPGE current, of about an order of magnitude larger than that observed in GaAs, InAs and SiGe low dimensional structures. The low symmetry of investigated samples opens a rich field for investigation of microscopic properties of this novel and promising material.

REFERENCES

[1] S.D. Ganichev and W. Prettl, *Intense Terahertz Excitation of Semiconductors*. Oxford University Press, 2006.